\documentclass[10pt,conference]{IEEEtran}
\usepackage{cite}
\usepackage{amsmath}
\usepackage{times}
\usepackage{latexsym}
\usepackage{amssymb}
\usepackage{setspace}
\usepackage{epsfig}
\usepackage{balance}
\usepackage{amssymb}
\usepackage[center]{caption2}
\usepackage{stfloats}
\usepackage{cases}
\usepackage{array}
\usepackage{pifont}
\usepackage{setspace}
\usepackage{fancyhdr}
\usepackage{amscd}
\usepackage{algorithm}
\usepackage{algorithmic}
\usepackage{pst-all}
\usepackage{pstricks}
\usepackage{epsfig}
\usepackage{balance}
\allowdisplaybreaks
\usepackage[T1]{fontenc}
\usepackage[latin1]{inputenc}

\allowdisplaybreaks

\def\bfmu{{\boldsymbol{\mu}}}

\IEEEoverridecommandlockouts
\makeatletter
  \def\vhrulefill#1{\leavevmode\leaders\hrule\@height#1\hfill \kern\z@}
\makeatother

\begin{document}

\title{Secure Multicast Communications with Private Jammers}
\author{\IEEEauthorblockN{Kanapathippillai Cumanan$^\dag$, Zhiguo Ding$^\ddag$,  Mai Xu$^\star$, and H. Vincent Poor$^\S$}
\IEEEauthorblockA{$^\dag$ Department of Electronics, University of York, York, YO10 5DD, UK.\\
$^\ddag$ School of Computing and Communications, Lancaster University, Lancaster, UK.\\
$^\star$ Department of Electrical Engineering, Beihang University, China.\\
$^\S$ Department of Electrical Engineering, Princeton University, Princeton, NJ, USA.}
}
\maketitle
\begin{abstract}
 This paper investigates secrecy rate optimization for a multicasting network, in which a transmitter broadcasts the same information to multiple legitimate users in the presence of multiple eavesdroppers. In order to improve the achievable secrecy rates, private jammers are employed to generate interference to confuse the eavesdroppers. These private jammers charge the legitimate transmitter for their jamming services based on the amount of interference received at the eavesdroppers. Therefore, this secrecy rate maximization problem is formulated as a Stackelberg game, in which the private jammers and the transmitter are the leaders and the follower of the game, respectively. A fixed interference price scenario is considered first, in which a closed-form solution is derived for the optimal amount of interference generated by the jammers  to maximize the revenue of the legitimate transmitter. Based on this solution, the Stackelberg equilibrium of the proposed game, at which both legitimate transmitter and the private jammers achieve their maximum revenues, is then derived.  Simulation results are also  provided to validate these theoretical derivations.
\end{abstract}
\IEEEpeerreviewmaketitle
\section{Introduction}
\indent The concept of information theoretic security   was first investigated in \cite{Wyner_J75} for wiretap channels by defining the concept of the secrecy capacity. Since then, information theoretic security has received considerable attention due to its low complexity   implementation and suitability for the dynamic   configurations of wireless networks, in which the physical layer characteristics of wireless channels  are exploited to establish secure communication between legitimate terminals. This novel paradigm complements the conventional cryptographic methods implemented  in the  upper networking layers by providing additional security at the physical layer.\\
\indent Multi-antenna terminals have the potential to enhance the performance of secret communications by exploiting spatial   degrees of freedom. However, the secrecy rates which are achievable by using  multi-antenna terminals are still limited by the quality of the wireless channels between the legitimate transmitter and the receivers, including the legitimate receivers and the eavesdroppers \cite{Ma_Sig_Process_J11,Gan_FD_Sig_Process_J13,Cuma_TVT_J14,Wei_Chen_Wireless_Lett_J15,Zheng_Wireless_Lett_J15,Zheng_IET_J15}. The existing works in \cite{Gan_FD_Sig_Process_J13,Ma_Sig_Process1_J13,Zheng_TVT_J15} demonstrate that  the performance of   secret communications can be further improved by using   cooperative jamming and artificial noise techniques, in which jamming signals are transmitted from external jammers or integrated  with the information bearing  signals sent by the legitimate transmitter. These approaches effectively degrade the capability of the eavesdroppers for retrieving  the legitimate users' signals, and hence enhance the achievable secrecy rates.\\
\indent Recently, game theoretic approaches have been applied   to the resource allocation problems in wireless  secret  communication networks \cite{Trappe_Securecom09,Liu_Info_security_J11,Poor_Wireless_Commun_J12,Basar_WiOpt09,Han_TVT_J12,Zhu_Han_ICMAS09,Swindlehurst_Sig_Process_J13,Swindlehurst_Sig_Process_J1_13,Zheng_EUSIPCO2014}. In \cite{Trappe_Securecom09}, a zero-sum game was formulated  for a secret communication  network by considering the signal-to-interference-plus-noise rate (SINR) difference between the legitimate receiver and the eavesdropper as the utility function.
The interaction among the nodes in cognitive radio networks has been investigated by using the Stackelberg game  \cite{Liu_Info_security_J11}. Cooperative game theory has been used  to improve the secrecy capacity of ad-hoc networks in \cite{Basar_WiOpt09}, and a distributed tree formation game was proposed for multihop wireless networks in \cite{Poor_Wireless_Commun_J12}. Physical layer security has been investigated   through a Stackelberg game for a two-way relaying network with unfriendly jammers in \cite{Han_TVT_J12},  and a distributed auction based approach has been used  to enhance the secrecy capacity in \cite{Zhu_Han_ICMAS09}. Jamming games have been formulated for multiple-input multiple-output (MIMO) wiretap channels with an active eavesdropper in \cite{Swindlehurst_Sig_Process_J13}, and a secrecy game for a Gaussian multiple-input single-output (MISO) interference channel has been investigated in \cite{Swindlehurst_Sig_Process_J1_13}.\\
\indent In this paper, a multicating network is considered as shown in Fig. \ref{fig:Multi_sec_network_jam}, where   all the legitimate users are to receive the same information in the presence of multiple eavesdroppers. In order to improve the achievable secrecy rates of the legitimate users, the private jammers are employed to generate artificial noise and confuse   the eavesdroppers. These private jammers introduce the costs for their jamming services based on the amount of interference generated  to the eavesdroppers. To compensate these jamming costs, the legitimate users   pay the transmitter for their enhanced secret  communications. Based on these interactions between the legitimate transceivers and  the private jammers, we formulate the   secrecy rate maximization problem as a Stackelberg game. A fixed interference price scenario is considered first and then a closed-form solution for the optimal amount of interference generated to each eavesdropper is obtained.   Based on this solution, we then investigate the corresponding Stackelberg equilibrium for the formulated game. In addition, simulation results are provided to validate the theoretical derivations of the proposed game theoretic approach.
\section{System Model}
\begin{figure}[t]
\includegraphics[scale = 0.7]{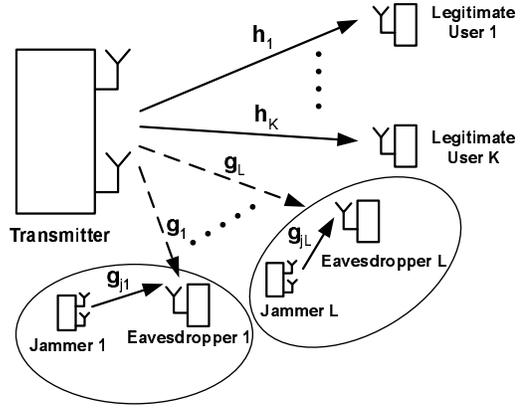}
\centering\caption{A multicasting secure  network with $K$ legitimate users, $L$ eavesdroppers and  $L$ private jammers.}\label{fig:Multi_sec_network_jam}

\end{figure}

\indent A secret communication  network with $K$ legitimate users, $L$ eavesdroppers and $L$ private jammers is considered in this paper,  as shown in Fig.\ref{fig:Multi_sec_network_jam}, where the transmitter broadcasts a common message to be received by all the legitimate users in the presence of multiple eavesdroppers. In this secure  network, the transmitter is equipped with  $N_{T}$ transmit antennas, whereas the legitimate users and the eavesdroppers are equipped with a single receive antenna, respectively. The channel coefficients between the legitimate transmitter and the $k^{\rm{th}}$ legitimate user as well as the $l^{\rm{th}}$ eavesdropper are denoted by $\mathbf{h}_{k} \in \mathbb{C}^{N_{T}\times 1}$ and $\mathbf{g}_{l} \in \mathbb{C}^{N_{T}\times 1}$, respectively.\\
\indent In addition, a set of private (friendly) jammers are employed to provide jamming services as shown in Fig.\ref{fig:Multi_sec_network_jam}. These private jammers generate artificial  interference to confuse  the eavesdroppers and they ensure that there is no interference leakage to the legitimate users. This is achieved by appropriately designing the beamformers at the jammers and employing a dedicated jammer near to each eavesdropper. Since, a dedicated jammer is closely located to the corresponding eavesdropper, each eavesdropper receives strong co-channel interference   from its corresponding private jammer.

Note that  these private jammers charge the legitimate transceivers  for their dedicated jamming services based on the amount of interference generated  to each eavesdropper. To compensate these interference costs, the legitimate transmitter  introduces the charges  to the legitimate users for their enhanced secure communications, by using the achievable  secrecy rates as the criteria.
The channel  gain between the $l^{\rm{th}}$ eavesdropper and the corresponding jammer is denoted  by $|g_{jl}|^{2}$. Furthermore, it is assumed that the legitimate transmitter and the jammers have the perfect channel state information of the eavesdroppers. This assumption is appropriate in a multicasting network, where potential eavesdroppers are also legitimate users of the network. This assumption has been commonly used in the literature \cite{Gan_CJ_Sig_Process_J11,Swindlehurst_Sig_Process_J11,Ma_Sig_Process1_J11}. The achievable secrecy rate at the $k^{\rm{th}}$   user can be written as follows:~\cite{Wornell_Info_Theory1_J10}
\begin{equation}\label{}
\small
    R_{k}\!\! =\!\!\left[\!\log\!\left(\!\!1\!+\!\frac{\mathbf{w}^{H}\mathbf{h}_{k}\mathbf{h}_{k}^{H}\mathbf{w}}{\sigma_{k}^{2}}\!\!\right)\!\!-\!\!\max_{1\leq l\leq L}\log\!\left(\!1\!+\!\frac{\mathbf{w}^{H}\mathbf{g}_{l}\mathbf{g}_{l}^{H}\mathbf{w}}{\sigma_{e}^{2}+p_{k}|g_{jk}|^{2}}\!\right)\!\right]^{+},
\end{equation}
where $\mathbf{w}\in \mathbb{C}^{N_{T}\times 1}$ and $p_{k}$ are the beamformer at the legitimate transmitter and the power allocation coefficient for the $k^{\rm{th}}$ private jammer, respectively. The $\sigma_{k}^{2}$ as well as $\sigma_{e}^{2}$ denote the noise variances at the $k^{\rm{th}}$ legitimate user and the eavesdropper, respectively, and $[x]^{+}$ represents $\max\{x,0\}$.
\section{Game Theoretic Approach for Secrecy Rate Optimization}
\indent In this section, we formulate the secrecy rate maximization problem into a Stackelberg game and then investigate the Stackelberg equilibrium for the proposed game. This game consists of two sets of players: a) leader and b) followers. All these players try to maximize their revenues, where the leaders first make a move and the followers will choose their strategies  according to the leaders' decisions. In the multicasting network considered in this paper, the private jammers (leaders) announce their interference prices   and then the legitimate transmitter (follower) determines the interference requirements according to the interference prices.\\
\indent The interference received at the $l^{\rm{th}}$ eavesdropper from the corresponding private jammer can be written as follows:
\begin{equation}\label{}
\small
    I_{l} = p_{l}|g_{jl}|^{2}.
\end{equation}
Here, we are only interested in the transmit power used by the jammer, where the beamformer at the jammer is appropriately designed to ensure that there is no interference leakage to the legitimate users. The private jammers aim to maximize their revenues by selling interference to the transmitter. The revenue of the $l^{\rm{th}}$ private jammer can be written as follows:
\begin{equation}\label{}
\small
    \phi_{l}(\mu_{l}) = \mu_{l}p_{l}|g_{jl}|^{2},
\end{equation}
where $\mu_{l}$ is the unit interference price charged by the corresponding jammer to cause interference at the $l^{\rm{th}}$ eavesdropper. Depending on the interference requirement at the $l^{\rm{th}}$ eavesdropper, the interference price should be determined by the corresponding jammer to maximize its revenue. These interference prices can be determined by solving  the following optimization  problem:
\begin{eqnarray}
\small
 \textrm{Problem (A):}~~~~~~ \max_{\bfmu\succeq \mathbf{0}}\!\!\!&&\!\!\!\sum_{l=1}^{L} \phi_{l}(\mu_{l},p_{l}),
\end{eqnarray}
where $\bfmu = [\mu_{1}\cdots \mu_{L}]$ includes the interference prices.\\
\indent On the other hand, the transmitter aims to maximize its revenue by charging the legitimate users based on their achievable  secrecy rates, where the revenue function at the transmitter can be written as follows:
\begin{equation}\label{}
\small
    \psi_{L}(\mathbf{p},\bfmu) = \sum_{k=1}^{K}\lambda_{k}R_{k}-\sum_{l=1}^{L}\mu_{l}p_{l}|g_{jl}|^{2},
\end{equation}
where $\lambda_{k}$ and $R_{k}$ are the unit price for the secrecy rate and the achievable  secrecy rate at the $k^{\rm{th}}$ user, respectively. It is assumed that the unit price   for each user is fixed at a predetermined  value. Hence, the transmitter should determine the beamforming vector as well as the interference requirements at different eavesdroppers in order to maximize its revenue. We first focus on the interference requirements at each eavesdropper with a fixed beamformer at the transmitter, which can be formulated into an optimization problem as follows:
\begin{eqnarray}
\small
  \textrm{Problem (B):}~~~~\max_{\mathbf{p}\succeq 0}\!\!\!&&\!\!\!\psi_{L}(\mathbf{p},\bfmu),
\end{eqnarray}
where $\mathbf{p} = [p_{1} \cdots p_{L}]$ represents the power allocation coefficients  at all jammers. \emph{Problem (A)} and  \emph{Problem (B)} form a Stackelberg game, and it is important to  investigate the corresponding Stackelberg equilibrium.
\subsection{Stackelberg Equilibrium}
\indent The Stackelberg equilibrium for the proposed game is defined as follows:\\
Stackelberg equilibrium: Let $\mathbf{p}^{*}$  be the optimal solution for \emph{Problem (B)} whereas $\bfmu^{*}$ contains the best prices for \emph{Problem (A)}. The solutions $\mathbf{p}^{*}$ and $\bfmu^{*}$ define the Stackelberg equilibrium point if the following conditions are satisfied for any set of $\mathbf{p}$ and $\bfmu$:
\begin{eqnarray}
\small
 \psi_{L}(\mathbf{p}^{*},\bfmu^{*})\!\!\!\!\!&\geq&\!\!\!\!\!\psi_{L}(\mathbf{p},\bfmu^{*}),~
   \phi_{l}(p_{l}^{*},\mu_{l}^{*})\!\geq\!\phi_{l}({p}_{l}^{*},\mu_{l}),\forall~l. \nonumber
\end{eqnarray}
\section{Stackelberg Equilibrium Solution}
\indent In this section, we derive the Stackelberg equilibrium solution for the proposed game. In order to analyze this equilibrium, the best response of the transmitter is first derived in terms of the interference requirement at each eavesdropper for fixed interference prices. Then, the optimal interference prices for the private jammers are obtained to maximize their revenues. These best responses can be derived by solving \emph{Problem (A)} and \emph{Problem (B)}. Particularly, we first solve the problem for a  scenario with fixed interference prices. Based on this solution, we then derive the Stackelberg equilibrium for the proposed game. Note that  we only consider the secure communication  network with a single legitimate user and multiple eavesdroppers. However, this can be easily extended for a scenario with multiple legitimate users and multiple eavesdroppers.
\subsection{Fixed Interference Prices}
\indent In this subsection, we focus on the fixed interference price scenario with a single legitimate user and multiple eavesdroppers. Note that for a particular user,     eavesdroppers with   large achievable  rates are more damaging since they significantly reduce the secrecy rate of this legitimate user. Therefore, introducing jamming to these   eavesdroppers will effectively  improve the achievable secrecy rate of the legitimate user.   Therefore, a set of eavesdroppers which have strong connections to the source are defined as super-active eavesdroppers. The rest of the eavesdroppers are referred as non-super-active eavesdroppers. The achievable secrecy rate of the legitimate user is defined as follows:
\begin{equation}\label{}
\small
    R_{1} = \log\!\left(1+\beta_{0}\right)\!-\!\max_{1\leq i\leq L}\log\left(\!1+\frac{\beta_{i}}{\sigma_{e}^{2}+p_{i}\alpha_{i}}\!\right),
\end{equation}
where
\begin{equation}\label{}
\small
    \beta_{0} = \frac{\mathbf{w}^{H}\mathbf{h}_{1}\mathbf{h}_{1}^{H}\mathbf{w}}{\sigma^{2}},~~\beta_{i} = \mathbf{w}^{H}\mathbf{g}_{i}\mathbf{g}_{i}^{H}\mathbf{w},~ \alpha_{i} = |g_{ji}|^{2}.
\end{equation}
The optimal interference requirements at each eavesdropper can be formulated as follows:
\begin{equation}\label{}
\small
\max_{\mathbf{p}\succeq\mathbf{0}}~\lambda_{1}R_{1}-\sum_{i\in\mathbb{K}}\mu_{i}p_{i}\alpha_{i},
\end{equation}
where vector $\mathbf{p}=[p_{1}\cdots p_{K}]$ includes the power allocation coefficients of the private jammers in the set $\mathbb{K}$ consisting of all super-active eavesdroppers. Without loss of generality, this problem can be reformulated as follows:
\begin{eqnarray}
  \max_{\mathbf{p}\succeq \mathbf{0},~t_{i},~t_{0}}\!\!\!\!\!&&\!\!\!\!\! \lambda_{1}\left[\log\!\left(1+\beta_{0}\right)-t_{0}\right]-\sum_{i=1}^{K}\mu_{i}p_{i}\alpha_{i}\nonumber\\
  \textrm{s.t.}\!\!\!\!\!&&\!\!\!\!\!\log\left(\!1+\frac{\beta_{i}}{\sigma_{e}^{2}+p_{i}\alpha_{i}}\!\right)\leq t_{i},~\forall~k\nonumber\\
  \!\!\!\!\!&&\!\!\!\!\!\max\{t_{1},\cdots,t_{K}\} = t_{0},~\forall~k,~t_{i}\geq 0,~\forall~k.\label{eq:SL-ME_Fint_1}
\end{eqnarray}
This problem is convex with  respect to  the power allocation coefficients at the private jammers and can be efficiently solved through interior point methods\cite{boyd_B04}.\\\\
\noindent\emph{Proposition 1}: By using the optimal solution of \eqref{eq:SL-ME_Fint_1}, the achievable  rates of the super-active eavesdroppers (i.e., $t_{i},~i\in\mathbb{K}$) will be equal and the power allocation coefficients  of the non-super-active eavesdroppers (i.e., $i\notin \mathbb{K}$) will be all zeros.\\\\
\noindent\emph{Proof}: Assume that $t_{i},~i\in\mathbb{K}$ are not equal. Assume that the minimum $t_{i}=t_{min}<t_{0}$ from all $t_{i},~i=1,\cdots,K$, and the corresponding $p_{i}$ will be higher than that of   $t_{min} = t_{0}$. Hence, the revenue of the transmitter (cost function of \eqref{eq:SL-ME_Fint_1}) with $t_{i}=t_{min}$ will be less than that with $t_{i}=t_{0}$. Thus, the achievable  rates of the super-active eavesdroppers (i.e., $t_{i},~i\in\mathbb{K}$) will be equal when  the optimal solution is used and the power allocation coefficients for  the non-super-active eavesdroppers (i.e., $i\notin \mathbb{K}$) will be zero.\hfill$\blacksquare$\\\\
Hence, the optimal interference requirements can be obtained by solving the convex problem in \eqref{eq:SL-ME_Fint_1}.
\subsection{Stackelberg Game}
\indent In this subsection, we formulate the problem into a Stackelberg game and investigate the Stackelberg equilibrium for the proposed game. In order to derive the equilibrium of the game, the best responses of   both the leaders and the follower should be obtained. The best response of the legitimate transmitter can be obtained by solving the following problem:
\begin{equation}\label{}
\max_{\mathbf{p}\succeq\mathbf{0}}~\lambda_{1}R_{SL-ME}-\sum_{i\in\mathbb{K}}\mu_{i}p_{i}\alpha_{i},
\end{equation}
where the vector $\mathbf{p}=[p_{1}\cdots p_{K}]$ consists of the power allocation coefficients of the private jammers in the super-active eavesdropper set $\mathbb{K}$. As we discussed in the previous subsection in \eqref{eq:SL-ME_Fint_1}, this problem is convex and the optimal power allocation can be obtained. Furthermore, the closed-form solution of this power allocation problem should be determined by deriving the Stackelberg equilibrium of the proposed game, as shown in the following lemma.\\\\
\emph{Lemma 1:} The optimal power allocation coefficient at the $i^{\rm{th}}$ jammer is given by
\begin{equation}\label{eq:power_allocation_SL-ME_close}
\small
    p_{i}^{*} = \frac{1}{\alpha_{i}}\left[\frac{\beta_{i}}{\gamma_{0}}-\sigma_{e}^{2}\right]^{+},
\end{equation}
where
\begin{eqnarray}
\small
   \beta_{i}\!\!&=&\!\!\mathbf{w}^{H}\mathbf{g}_{i}\mathbf{g}_{i}^{H}\mathbf{w}\nonumber\\
   \gamma_{0}^{*}\!\! \!\!&=&\!\!\frac{\sum_{i=1}^{K}\mu_{i}\beta_{i}\!+\!\sqrt{\sum_{i=1}^{K}\!\mu_{i}\beta_{i}\left(4\lambda_{1}\!+\!\sum_{i=1}^{K}\mu_{i}\beta_{i}\right)}}{2\lambda_{1}}
\end{eqnarray}
\noindent\emph{Proof:} Please refer to Appendix A.\hfill$\blacksquare$\\\\
\indent The private jammers need to announce their interference prices to maximize their revenues. These optimal interference prices can be obtained by solving the following problem:
\begin{equation}\label{}
\small
    \max_{\bfmu\succeq \mathbf{0}} \sum_{l=1}^{L}\phi_{i}(p_{i}^{*},\mu_{i}) = \sum_{l=1}^{L}\mu_{i}p_{i}^{*}\alpha_{i}.
\end{equation}
Based on the closed-form solution of the optimal power allocation coefficients $p_{i}^{*}$s in \eqref{eq:power_allocation_SL-ME_close} in terms of the interference prices $\mu_{i}$s, the optimal interference prices problem can be reformulated as
\begin{equation}\label{eq:Rev_jamm_SL_ME}
\small
    \max_{\bfmu\succeq \mathbf{0}} \frac{2\lambda_{1}\sum_{i=1}^{K}\mu_{i}\beta_{i}}{\sum_{i=1}^{K}\mu_{i}\beta_{i}\!\!+\!\!
    \sqrt{\sum_{i=1}^{K}\!\mu_{i}\beta_{i}\left(\!4\lambda_{1}\!+\!\sum_{i=1}^{K}\mu_{i}\beta_{i}\!\right)}}-\sigma_{e}^{2}\sum_{i=1}^{K}\mu_{i}.
\end{equation}
The optimal interference prices $\mu_{i}$s can be obtained by solving the above problem through existing numerical methods. However, the closed-form solutions of these interference prices are not easy to derive. Therefore, we assume the use of  the same interference price (uniform interference price) for all private jammers (i.e., $\mu_{1} = \mu_{2} = \cdots = \mu_{K}= \mu_{0}$). The problem in \eqref{eq:Rev_jamm_SL_ME} can be formulated with the uniform interference price as follows:
\begin{equation}\label{eq:Rev_jamm_SL_ME1}
\small
    \max_{\mu_{0}\geq 0} \frac{2\lambda_{1}\mu_{0}\sum_{i=1}^{K}\beta_{i}}{\mu_{0}\sum_{i=1}^{K}\beta_{i}\!\!+\!\!\sqrt{\mu_{0}\sum_{i=1}^{K}\!\beta_{i}\left(\!4\lambda_{1}\!+\mu_{0}\!\sum_{i=1}^{K}\beta_{i}\!\right)}}-K\sigma_{e}^{2}\mu_{0}.
\end{equation}
\emph{Lemma 2:} The optimal interference price $\mu_{0}^{*}$ in \eqref{eq:Rev_jamm_SL_ME1} is given by
\begin{equation}\label{}
\small
    \mu_{0}^{*} = \frac{0.5\left[-4\lambda_{1}K\sigma^{2}\eta_{1}+2\lambda_{1}\sqrt{K\sigma^{2}\eta_{2}+4K^{2}\sigma^{4}\eta_{1}^{2}}\right]}{K\sigma^{2}\eta_{2}}
\end{equation}
where
\begin{equation}
\small
  \eta_{1}=\left(1+\frac{K\sigma^{2}}{\bar{c}_{2}}\right),~\eta_{2}=\left(\bar{c}_{2}+K\sigma^{2}\right),~\bar{c}_{2} = \sum_{i=1}^{K}\beta_{i}.
\end{equation}
\noindent\emph{Proof:} Please refer to Appendix B.\hfill$\blacksquare$\\\\
Hence, the Stackelberg equilibrium of the proposed game with uniform interference price can be defined by ($p_{i}^{*}~\forall~ i,\mu_{0}^{*}$), at which both the transmitter and the private jammers maximize their revenues.
\section{Simulation Results}
\indent In this section, we validate the derived theoretical results by using computer simulations. Here, we consider a multicasting network with a single legitimate user and  two eavesdroppers, where the transmitter broadcasts the same information to all the legitimate users in the presence of multiple eavesdroppers. In addition, private jammers are employed to confuse the eavesdroppers by introducing interference, which will improve the achievable secrecy rates of the legitimate users. It is assumed that the legitimate transmitter is equipped with three antennas whereas the legitimate user and  the eavesdroppers have a single antenna. The channel coefficients between all the terminals are generated through zero-mean circularly symmetric independent and identically distributed complex Gaussian random variables and the noise variance at all the terminals is assumed to be 0.1. In the following subsections, we provide simulation results for the scenario  with the fixed interference prices and the Stackelberg game scenario, respectively.
\subsection{Fixed Interference Prices}
\indent In this subsection, we evaluate the performance of the proposed schemes with fixed interference prices at the private jammers. The fixed unit interference prices at the jammers are assumed to be 1 and 3 (i.e., $\mu_{1}=1,~\mu_{2}=3$), respectively. Table 1 provides the theoretical and simulation based optimal power allocation coefficients and the corresponding revenues of the legitimate transmitter for different sets of channels. These results validate the derivation of the theoretical results which are indistinguishable with the simulation based results.
\begin{table*}[t]
\small
  \centering
\begin{tabular}{|c|c|c|c|c|c|c|c|c|}
  \hline
  Channels & \multicolumn{2}{c|}{\begin{tabular}{c}
                                     Power Allocation:\\
                                    Jammer 1 \\
                                  \end{tabular}} & \multicolumn{2}{c|}{\begin{tabular}{c}
                                     Power Allocation:\\
                                    Jammer 2 \\
                                  \end{tabular}} & \multicolumn{2}{c|}{\begin{tabular}{c}
                                     Achieved\\
                                    Secrecy Rate \\
                                  \end{tabular}} & \multicolumn{2}{c|}{\begin{tabular}{c}
                                     Revenue:\\
                                    Legitimate Transmitter\\
                                  \end{tabular}}  \\
  \cline{2-9}
   & Derivation & Simulation & Derivation & Simulation & Derivation & Simulation & Derivation & Simulation\\
   \hline
 Channel 1 & 0.3324  & 0.3324 & 0.7457  & 0.7458 & 2.7083 & 2.7241 & 13.0855 & 12.8145 \\
  \hline
  Channel 2 & 0.1264  & 0.1264 & 0.5729  & 0.5430 & 3.3334 & 3.3223 & 15.2002 & 15.2016 \\
  \hline
  Channel 3 & 3.3886  & 3.3889 & 1.0284  & 1.0284 & 2.8085 & 2.8234 & 13.4161 & 13.4203 \\
  \hline
 Channel 4 & 1.1613  & 1.1614 & 1.0441  & 1.0442 & 2.9185 & 2.9296 & 13.7907 & 13.7928 \\
  \hline
  Channel 5 & 0.2778  & 0.2778 & 2.0209  & 2.0211 & 3.2938 & 3.2949 & 15.1031 & 15.1031 \\
  \hline
\end{tabular}
\label{Table:FixintSLME}
\renewcommand\thetable{1}
\caption{The optimal power allocation of the private jammers with fixed interference prices $\mu_{1} =1$ and $\mu_{2} =3$, achievable  secrecy rates and revenues of legitimate transmitter obtained from closed-form solution and simulation for different sets of channels. The unit price for the achievable  secrecy rate at the legitimate user is 5 ($\lambda_{1}=5$).}
\end{table*}
\begin{table*}[t]
\small
  \centering
\begin{tabular}{|c|c|c|c|c|c|}
  \hline
  Channels & \multicolumn{2}{c|}{
                                     Interference Price:
                                  } & \multicolumn{2}{c|}{
                                     Revenue of Jammers:
                                  } & \begin{tabular}{c}
                                     Stackelberg Equilibrium:\\
                                    $\left(p_{1}^{*},p_{2}^{*},\mu_{0}^{*}\right)$ \\
                                  \end{tabular}  \\
  \cline{2-5}
   & Derivation & Simulation & Derivation & Simulation & \\
  \hline
 Channel 1 & 4.0721  & 4.1000 & 1.5381  & 1.5378 & $\left(0.0677,0.3070,4.0721\right)$ \\
  \hline
 Channel 2 & 2.1647  & 2.2000 & 0.5372  & 0.5378 &  $\left(0.3076,0.6900,2.1647\right)$\\
  \hline
 Channel 3 & 2.6639  & 2.7000 & 0.7088  & 0.7084 &  $\left(0.1501,1.0917,2.6639\right)$ \\
  \hline
 Channel 4 & 3.1023  & 3.1000 & 0.8887  & 0.8892 &  $\left(0.1501,0.6996,3.1023\right)$\\
  \hline
  Channel 5 & 4.0322  & 4.0000 & 1.4932  & 1.4935 &  $\left(2.5895,0.7858,4.0322\right)$\\
  \hline
\end{tabular}
\label{Table:Stack_game2}
\renewcommand\thetable{2}
\caption{The optimal interference prices and revenues of the private jammers as well as Stackelberg equilibrium  for different sets of channels. The unit price for the achievable  secrecy rate at the legitimate user is 5 ($\lambda_{1}=5$).}
\end{table*}
\subsection{Stackelberg Game}
\indent In this subsection, we validate the derived Stackelberg equilibrium of the proposed game. Table 2 provides the derived theoretical  Stackelberg equilibrium and the  simulation based one,  as well as the corresponding jammer revenues with the uniform interference price assumption  (i.e., $\mu_{1} = \mu_{2} = \mu_{0}$) for different sets of channels. The simulation  results   are consistent to the theoretical ones and validate the Stackelberg equilibrium of the proposed game for different sets of channels. It is worth pointing out that any deviations  from these equilibria caused by different strategies of the legitimate transmitter and the jammers will introduce loss in their revenues.
\section{Conclusions}
\indent In this paper, we studied the secrecy rate optimization problem for a multicasting network, where   multiple users are to receive the same information in the presence of multiple eavesdroppers. To improve the secrecy rate performance, private jammers were employed to generate  interference to the eavesdroppers. In addition, these jammers charge the legitimate transceivers for their jamming services. This optimization  problem was formulated into a Stackelberg game, where the private jammers and the legitimate transmitter are the players of the game. We first focused on the fixed interference price scenario and a closed-form solution was derived for the optimal interference requirements. Based on this solution, a Stackelberg equilibrium was derived to maximize the revenues of both the legitimate transmitter and the private jammers. Simulation results were provided to support the derived theoretical results.
\begin{figure*}
\small{
\begin{eqnarray}\label{eq:diff_jamm_revenue}
\frac{\partial f(\mu_{0})}{\partial\mu_{0}} &=&\frac{2\lambda_{1}\bar{c}_{1}}{\mu_{0}\bar{c}_{1}+q}-\frac{2\lambda_{1}\bar{c}_{1}\mu_{0}\left(\bar{c}_{1}+\frac{\bar{c}_{1}^{2}\mu_{0}+2\lambda_{1}\bar{c}_{1}}{\mu_{0}\bar{c}_{1}+q}\right)}{\left(\mu_{0}\bar{c}_{1}+q\right)^{2}},~\textrm{where}~q=\sqrt{\mu_{0}\bar{c}_{1}(4\lambda_{1}+\mu_{0}\bar{c}_{1})},~~\bar{c}_{1} = \sum_{i=1}^{K}\beta_{i}\nonumber\\
\frac{\partial^{2}f(\mu_{0})}{\partial\mu_{0}^{2}}&=& \frac{-4\lambda_{1}\bar{c}_{1}\left(\bar{c}_{1}+\frac{\bar{c}_{1}^{2}\mu_{0}+2\lambda_{1}\bar{c}_{1}}{q}\right)}{\left(\bar{c}_{1}\mu_{0}+q\right)^{2}}+\frac{4\lambda_{1}\bar{c}_{1}\mu_{0}\left(\bar{c}_{1}+\frac{\bar{c}_{1}^{2}\mu_{0}+2\lambda_{1}\bar{c}_{1}}{q}\right)^{2}}{\left(\bar{c}_{1}\mu_{0}+q\right)^{3}}-\frac{2\lambda_{1}\bar{c}_{1}\mu_{0}\left(\frac{\bar{c}_{1}^{2}}{q}-\frac{\left(\bar{c}_{1}\mu_{0}+2\lambda_{1}\bar{c}_{1}\right)^{2}}{q^{3}}\right)}{\left(\bar{c}_{1}\mu_{0}+q\right)^{2}}
\end{eqnarray}}
\hrule
\end{figure*}
\begin{figure*}
\small{
\begin{eqnarray}\label{eq:second_diff_jamm_revenue1}
\frac{\partial^{2}f(\mu_{0})}{\partial\mu_{0}^{2}}\!\!\!\!\!\!&=&\!\!\!\!\!\! \frac{-4\lambda_{1}\bar{c}_{1}^{2}q(q+\bar{c}_{1}\mu_{0}+2\lambda_{1})[q^{2}-\bar{c}_{1}\mu_{0}(\bar{c}_{1}\mu_{0}+\lambda_{1})]-2\lambda_{1}\bar{c}_{1}^{3}\mu_{0}\left(\bar{c}_{1}\mu_{0}+q\right)\left[q^{2}-(\bar{c}_{1}\mu_{0}+2\lambda_{1})^{2}\right]}{q^{3}\left(\bar{c}_{1}\mu_{0}+q\right)^{3}}
\end{eqnarray}}
\begin{eqnarray}\label{eq:second_diff_jamm_revenue2}
\small
\textrm{By substituting}~~q &=&\sqrt{\mu_{0}\bar{c}_{1}(4\lambda_{1}+\mu_{0}\bar{c}_{1})},\Longrightarrow \frac{\partial^{2}f(\mu_{0})}{\partial\mu_{0}^{2}} = \frac{-12\lambda_{1}^{2}\bar{c}_{1}^{3}q\mu_{0}\left(q+\bar{c}_{1}\mu_{0}+2\lambda_{1}\right)-8\lambda_{1}^{3}\bar{c}_{1}^{3}\mu_{0}\left(\bar{c}_{1}\mu_{0}+q\right)}{q^{3}\left(\bar{c}_{1}\mu_{0}+q\right)^{3}}< 0
\end{eqnarray}
\hrule
\end{figure*}
\section*{Appendix A: proof of Lemma 1}
With the optimal power allocation coefficients in \eqref{eq:SL-ME_Fint_1}, the achievable  rates of the super-active eavesdroppers (i.e., $i \in \mathbb{K} $) will be equal as stated in \emph{Proposition 1}. Hence, the power allocation coefficient at the $i^{\rm{th}}$ private jammer can be written as follows:
\begin{equation}\label{}
\small
    \frac{\beta_{i}}{\sigma_{e}^{2}+p_{i}\alpha_{i}}=\gamma_{0},\Longrightarrow p_{i} = \frac{1}{\alpha{i}}\left[\frac{\beta_{i}}{\gamma_{0}}-\sigma_{e}^{2}\right]^{+}.
\end{equation}
The original optimization problem in \eqref{eq:SL-ME_Fint_1} can be formulated in terms of $\gamma_{0}$ as follows:
\begin{eqnarray}\label{eq:revenue_LT_gamma0_game}
\small
   \max_{\gamma_{0}\geq 0}\!\!\!\!\!&&\!\!\!\!\lambda_{1}\left[\log(1+\beta_{0})\!-\!\log(1\!+\!\gamma_{0})\right]\!\!-\!\! \frac{1}{\gamma_{0}}\sum_{i=1}^{K}\mu_{i}\beta_{i}\!+\!\sigma_{e}^{2}\sum_{i=1}^{K}\mu_{i}\nonumber\\
   \!\!\!\!\!&\triangleq&\!\!\!\!\!f(\gamma_{0})
\end{eqnarray}
The optimal $\gamma_{0}^{*}$ should satisfy the KKT conditions and therefore we obtain the following:
\begin{eqnarray}
\small
   \frac{\partial f(\gamma_{0})}{\partial \gamma_{0}}\!=\! -\!\frac{\lambda_{1}}{1\!+\!\gamma_{0}}\!+\!\frac{\tau}{\gamma_{0}^{2}},~\frac{\partial^{2} f(\gamma_{0})}{\partial \gamma_{0}^{2}}\!=\! \frac{\lambda_{1}}{(\!1\!+\!\gamma_{0}\!)^{2}}\!-\!\frac{2\tau}{\gamma_{0}^{3}},
\end{eqnarray}
where $\tau = \sum_{i=1}^{K}\!\mu_{i}\beta_{i}$. The function $f(\gamma_{0})$ is concave if the following condition is satisfied:
\begin{equation}\label{}
\small
    \frac{\gamma_{0}^{3}}{(1+\gamma_{0})^{2}}\leq \frac{2\tau}{\lambda_{1}}.
\end{equation}
Hence, the optimal $\gamma_{0}^{*}$ can be obtained if $\lambda_{1}$ is large enough to satisfy the above condition. This means that the legitimate transmitter should charge the legitimate user a reasonable price to make a profit. Note that  the optimal $\gamma_{0}^{*}$ should satisfy the KKT conditions.
\begin{equation}\label{}
\small
    \frac{\partial f(\gamma_{0})}{\partial \gamma_{0}}=0.
\end{equation}
The optimal $\gamma_{0}^{*}$ can be obtained by solving the following equation:
\begin{equation}\label{}
\small
    \lambda_{1}\gamma_{0}^{2}-\gamma_{0}\sum_{i=1}^{K}\mu_{i}\beta{i}-\sum_{i=1}^{K}\mu_{i}\beta{i}= 0.
\end{equation}
and $\gamma_{0}>0$,
\begin{equation}
\small
\gamma_{0}^{*} = \frac{\sum_{i=1}^{K}\mu_{i}\beta_{i}+\sqrt{\sum_{i=1}^{K}\mu_{i}\beta_{i}\left(4\lambda_{0}\!+\!\sum_{i=1}^{K}\mu_{i}\beta_{i}\right)}}{2\lambda_{1}}.
\end{equation}
Hence the optimal power allocation coefficient of the $i^{\rm{th}}$ can be written as follows:
\begin{equation}\label{}
\small
    p_{i}^{*} = \frac{1}{\alpha{i}}\left[\frac{\beta_{i}}{\gamma_{0}^{*}}-\sigma_{e}^{2}\right]^{+}.
\end{equation}
This completes the proof of \emph{Lemma 1}.\hfill$\blacksquare$
\section*{Appendix B: proof of Lemma 2}
\indent We first show that the revenue function of the jammers in \eqref{eq:Rev_jamm_SL_ME1} is  concave in terms of $\mu_{0}$ for $p_{i}>(0)$ in \eqref{eq:power_allocation_SL-ME_close},  and then we derive the optimal interference price $\mu_{0}^{*}$. The revenue function of the jammers is defined as follows:
\begin{equation}\label{eq:rev_jamm_appndx_SL_ME}
\small
    f(\mu_{0}) = \frac{2\lambda_{1}\mu_{0}\bar{c}_{1}}{\mu_{0}\bar{c}_{1}\!\!+\!\!\sqrt{\mu_{0}\bar{c}_{1}\left(\!4\lambda_{1}\!+\mu_{0}\!\bar{c}_{1}\!\right)}}-K\sigma_{e}^{2}\mu_{0},
\end{equation}
where $\bar{c}_{1} = \sum_{i=1}^{K}\beta_{i}$. The concavity of $f(\mu_{0})$ can be proven by finding the second derivative with respect to $\mu_{0}$ as in \eqref{eq:diff_jamm_revenue}. In order to prove that the function in \eqref{eq:rev_jamm_appndx_SL_ME} is concave, we need to show that the second derivative (i.e.,$\frac{\partial^{2}f(\mu_{0})}{\partial\mu_{0}^{2}}$) is negative. This has been proved in \eqref{eq:second_diff_jamm_revenue1} and \eqref{eq:second_diff_jamm_revenue2} which are in the previous page. This confirms that the revenue function of the jammers is concave in $\mu_{0}$ and the optimal $\mu_{0}^{*}$ should satisfy the KKT conditions $\frac{\partial f(\mu_{0})}{\partial\mu_{0}}=0$ \cite{boyd_B04}:
\begin{eqnarray}
\small
     \frac{2\lambda_{1}\bar{c}_{1}}{\mu_{0}\bar{c}_{1}+q}-\frac{2\lambda_{1}\bar{c}_{1}\mu_{0}\left(\bar{c}_{1}+\frac{\bar{c}_{1}^{2}\mu_{0}+2\lambda_{1}\bar{c}_{1}}{\mu_{0}\bar{c}_{1}+q}\right)}{\left(\mu_{0}\bar{c}_{1}+q\right)^{2}}  &=& 0,
\end{eqnarray}
{\small
\begin{equation}
    \mu_{0}^{*} = \frac{0.5\left[-4\lambda_{1}K\sigma^{2}\eta_{1}+2\lambda_{1}\sqrt{K\sigma^{2}\eta_{2}+4K^{2}\sigma^{4}\eta_{1}^{2}}\right]}{K\sigma^{2}\eta_{2}}.\nonumber
\end{equation}}
This completes the proof of \emph{Lemma 2}.\hfill$\blacksquare$
\balance
\bibliographystyle{ieeetr}
\bibliography{Myjournals,myrefs,mybooks}
\end{document}